\documentclass{svproc}
\usepackage{graphicx}
\usepackage{multicol}
\usepackage{footmisc}

\usepackage{url}

\begin{document}
\mainmatter              
\title{User Perception and Actions Through Risk Analysis Concerning Cookies}
\titlerunning{User Risk Perception About Concerning Cookies} 
\author{Matthew Wheeler \and Suleiman Saka \and
Sanchari Das}
\authorrunning{Wheeler et al.} 
\tocauthor{Matthew Wheeler, Suleiman Saka, Sanchari Das}
\institute{University of Denver, Colorado C0 80208, USA.\\
\email\{{matthew.wheeler}, \email{suleiman.saka}, \email{sanchari.das}\}{@du.edu}
}

\maketitle            

\begin{abstract}

A \textit{website browser cookie} is a small file created by a web server upon visitation, which is placed in the user's browser directory to enhance the user's experience. However, first and third-party cookies have become a significant threat to users' privacy due to their data collection methods. To understand the users' perception of the risk of cookies and targeted advertisements, we conducted a user study through a control versus experimental group survey. Our goal was to gauge how user knowledge reflected their security and privacy preferences on the internet; thus, for the experimental group, we created a learning website and information videos through participatory design in a workshop with $15$ participants. After that, by evaluating the responses of $68$ participants through the survey, we analyzed user awareness of cookies, their privacy implications, and how risk communication can impact user behavior. 

\keywords{Cookies, Third-party Data Collection, Privacy Concerns, Risk Analysis, Targeted Advertisements.}
\end{abstract}
\section{Introduction}
With the progression into the digital era, users have become increasingly dependent on websites to carry out their daily activities, and an overwhelming amount of websites now use web cookies. Browser cookies or website cookies are small files created by websites' servers to enhance user experience with the website usage~\cite{sanchez-rola2019can}. There are two primary browser cookies; first-party cookies and third-party cookies\cite{chen2021cookie}. 
Given the nature of the cookies, the files are often used to save features like login credentials, user preferences, and lists saved by users for future references, such as virtual shopping carts, among other things~\cite{demir2022towards}. 

Thus, evaluating how users perceive digital risks is critical as to whether risk communication can help inform users of these threats. We conducted a workshop where we designed risk communication videos with $15$ participants through participatory design. After that, we used this video to perform a control versus experimental group survey. We collected data from $68$ participants to study users' browsing habits, their knowledge of websites' cookies, and how they interact with them. The survey results indicate that most participants needed more information about cookies and their uses. In contrast, most have a good computer and security background. 

\section{Related Work}
\label{sec:related}
Websites use cookies to save users' information, such as; registered passwords, usernames, credit card numbers, language selections, etc.~\cite{marino2021privacy}, so the web server can read this data when the user logs back into the website again. Over the years, there has been extensive research on cookies and security issues that have been addressed.
\subsection{First-Party and Third-Party Cookies}
The first-party cookies (also called persistent cookies) are responsible for collecting and storing users' previous interactions with the website~\cite{egelman2013my,hu2021cccc,chen2021cookie}. This type of cookie is not deleted until the time-out value expires, even if the browsing session ends~\cite{ayeson2011flash,cahn2016empirical,kesmodel2005cookies}. Third-party cookies aim to study the users' behavior through their choices and preferences across multiple websites and use this personal information for advertising, social media, and other web applications~\cite{pierson2011social,oksanen2022companies}. Third-party cookies are created to discover and evaluate users' online footprint on a large scale~\cite{hu2021cccc,ramlakhan2011ethical,geradin2020google}. 

\subsection{Privacy Concern}
Collecting accurate information about the user and analyzing it has become a threat physically and psychologically to the individual's privacy\cite{mittal2010user}. Moreover, companies can turn this free information into millions of dollars through advertising designed to influence the individual\cite{agarwal2020stop,li2015trackadvisor,smit2014understanding,korac2020information}. In addition, a significant amount of cookies are collected from unknown or untrusted parties. For example, 44.08\% of cookies are gathered for parties that are registered not to exist~\cite{li2015trackadvisor,hu2021cccc,mayer2012third}. However, companies like Microsoft use 72.2\% of third-party cookies in their websites and browsers, and Google uses 25\% of third-party tracking directors of certain popular websites~\cite{li2015trackadvisor,milne2009toward}. 
\subsection{User Awareness}
Flinn and Lumsden found in their studies that many users have tried to teach themselves to protect their online privacy and security~\cite{flinn2005user}. Privacy awareness allows users to build a firewall to defend themselves from privacy risks and maintain control of their online footprint\cite{pelau2020consumers'}. Furthermore, making users aware of third-party cookies and their uses, risks, and benefits will help reduce the widespread sharing of people's information~\cite{boerboom2020cambridge,boerman2021exploring}. Awareness can also help to find and reduce overall risks using risk communication~\cite{Das2020User,das2019all}. Our research aims to determine whether people's knowledge of the risks of third parties' and websites' cookies will raise the degree of their protection towards the privacy of their information and thus reduce the acceptance of the use of cookies. 
\section{Method}
\label{sec:method}
Our study aims to measure users' concerns and awareness about websites' cookies. Developing a participatory design to evaluate the risk communication with a short video was critical to understanding how it impacts the general idea of website cookies' danger to participants. So, the participatory design incorporated other informative ideas in the video necessary to aid participants' knowledge. This was initially tested with $15$ participants before deploying it later to both the control and experimental group user study. This survey on $68$ participants was launched to evaluate the users' perception based on specific metrics initially set out by our research. The in-lab survey also guaranteed participants' opinions reflect their informative support and how risk communication plays a vital role in their decision-making.

\subsection{Creating An Educational Website}
We implemented a mixed methods approach to create the educational website to tap into the risk communication and evaluate its impact. The first part was conducting a workshop that included the concept of participatory design to develop the website. We invited $15$ participants to the workshop who were first interviewed to ask about their perception of cookies; after that, we discussed how educational information could be communicated to the users. We got mixed responses, both textual ($4$), video-based risk communication ($5$), and both ($6$); thus, we implemented both strategies~\cite{das2020smart} for educating the users. We designed the website to make it simple for the user to navigate and understand cookies. We focused on having the critical information visible on the home page and improved upon it as per the participants' feedback. As for the appearance, we wanted our website to be accessible and comfortable, in calm colors and proportion to our subject. Initially, we used instructive videos with illustrations to explain how the cookies worked. But, based on the users' feedback, we used screen recording software to make the video that explains our website. After showing the information, we asked the participants to try using the \textit{www.cookieserve.com} website as a hands-on experience to determine the number of cookies on their favorite websites.

\subsection{User Study: Control vs. Experimental Group - Survey}
We divided our survey into several categories. The first section focused on determining users' browser preferences and their current practices with blocking or monitoring cookies. These questions would give us a baseline for current practices, tell us if web browser preference played a part in users' online security practices, and help us to understand the users' baseline perception of web cookie usage. This section was followed by questions to determine the users' cyber background, helping us know how technological knowledge affected web browsing practices. Regarding participants' understanding of cookies is where our survey differed between the experimental and control groups. The questions are regarding the participant's current knowledge of cookies. For the \textit{control group}, it will give us a glimpse into the general public's knowledge of cookies. For the \textit{experimental group}, it will tell us if the website and video were effective in informing participants about cookies. 
 
Following the cookie knowledge questions, we asked about users' perceptions of cookie usage. This section was created to see how cookie perception and action differ when users are informed about cookies versus when users are not informed. We followed these questions with ones to gauge how dangerous feel cookies are to their online privacy and how they feel the need to mitigate cookie collection. We wanted to see how responses differed among informed and uninformed users. In the last section, we asked participants about the monitoring and blocking websites' cookies tools to know if they are using one and their opinion on it. In the end, we finished the survey by setting demographic questions such as education level, ethnicity, and additional question if the participants wanted to share something with us. Between the sections, we added attention questions to get accurate data.

\subsection{Thematic Analysis}
To ensure the integrity of our data, we excluded all participants' surveys who did not qualify to participate (did not agree to participate, were under 18, were not a US resident, did not agree to answer honestly and to their best knowledge, or those who incorrectly answered the attention questions). After the initial thematic analysis, we got $68$ participants who qualified for the study. We then categorized participants' answers by age and educational level to explore how these factors resulted in a difference in technology and web safety awareness. Finally, we focused on the participant's perception of cookies and their actions to monitor and block them. The data was analyzed and coded by two researchers involved in the study. We had an inter-rater reliability of 89.7\% for the survey. For the workshop and the participatory design, we developed the website based on the input of all the participants. 
 
\section{Results}
\label{sec:results}
The study limited the participants to those currently residing in the US and are above $18$ years old. The average age of participants was between $18-24$ years old, and we had some participants between the age of $51-60$. The participants are volunteers of different educational levels 15.87\% high school graduates, 41.27\% bachelor's degree program, 20.63\% master's degree program, 11.11\% Doctorate, and 11.12\% others. The proportion of males was 50.79\%, females 44.44\%, and do not wish to specify 4.76\%. Table~\ref{tab:1} details the demographic background of the participants.

\begin{table}
    \centering
    \caption{The demographic of participants that we use in analyzing}
    \label{tab:1}
    \begin{tabular}{cccccc}
     \hline
        \textbf{Age} & \textbf{No. of P}& \textbf{Education Level} & \textbf{No. of P}&\textbf{Ethnicity} & \textbf{No. of P} \\
     \hline
         18-24 years &24&Less than high school&1&Asian& 11 \\
         \hline
         25-30 years& 12&      High school graduate&   10&     African-American&           2 \\
         \hline
         31-40 years& 12&      Diploma&                4&      Latino or Hispanic&         5    \\
         \hline
         41-50 years& 6&      Bachelors degree&       26&     White&                      33\\
         \hline
         51-60 years& 9&      Masters degree&         13&     Arab&                       3  \\
         \hline
         
         61-70 years& 5&      Doctorate&              7&  Two or more   &      2\\
         
           \hline
        -& -&     Don't wish to say &         1&     Don't wish to say &                        6\\
          \hline
         
        -& -&     Professional Degree &         1&     - & -  \\
        \hline 
         
    \end{tabular}
     
\end{table}

Firstly, we aimed to study the participants' computer and security backgrounds. Therefore, we asked participants five questions about how strongly they would consider them-self good in technology and able to control the security risk. Consequently, most participants had relatively good technical and security backgrounds. However, the most significant percentage were those who felt they did not have the confidence to protect themselves from security risks. Before we delved more deeply into cookies, we needed to study the participants' behaviors and how they interact with cookies and browsing in general. First, 73.53\% of participants prefer using Google Chrome for the following reasons: easy to use, the default browser on their devices, and a large portion of them mentioned that they like the password auto-fill feature and history visits of websites. Moreover, most participants believe that Google Chrome, Mozilla Firefox, DuckDuckGo, and Safari browsers can protect their privacy. 

About 64.62\% of participants indicated their desire to make their locations unknown to the applications and websites. Also, they want to block cookies used to create target ads. Nevertheless, 36.92\% of participants want to avoid blocking cookies and do not care to prevent sharing their location. Regarding using tools to protect participants' devices from third parties, 49.21\% of participants need the means to block and monitor cookies. At the same time, about 68.19\% use AdBlock Plus, uBlock Origin, and Privacy Badger. When we asked about the reason for those who did not have any tools to block or monitor their cookies, most answered that they did not know any means, and others said they needed to understand why they needed one and how to use it.

\subsection{Comparison Between Control and Experimental Groups}
 \begin{figure}
\centering
\includegraphics[width=0.7\textwidth]{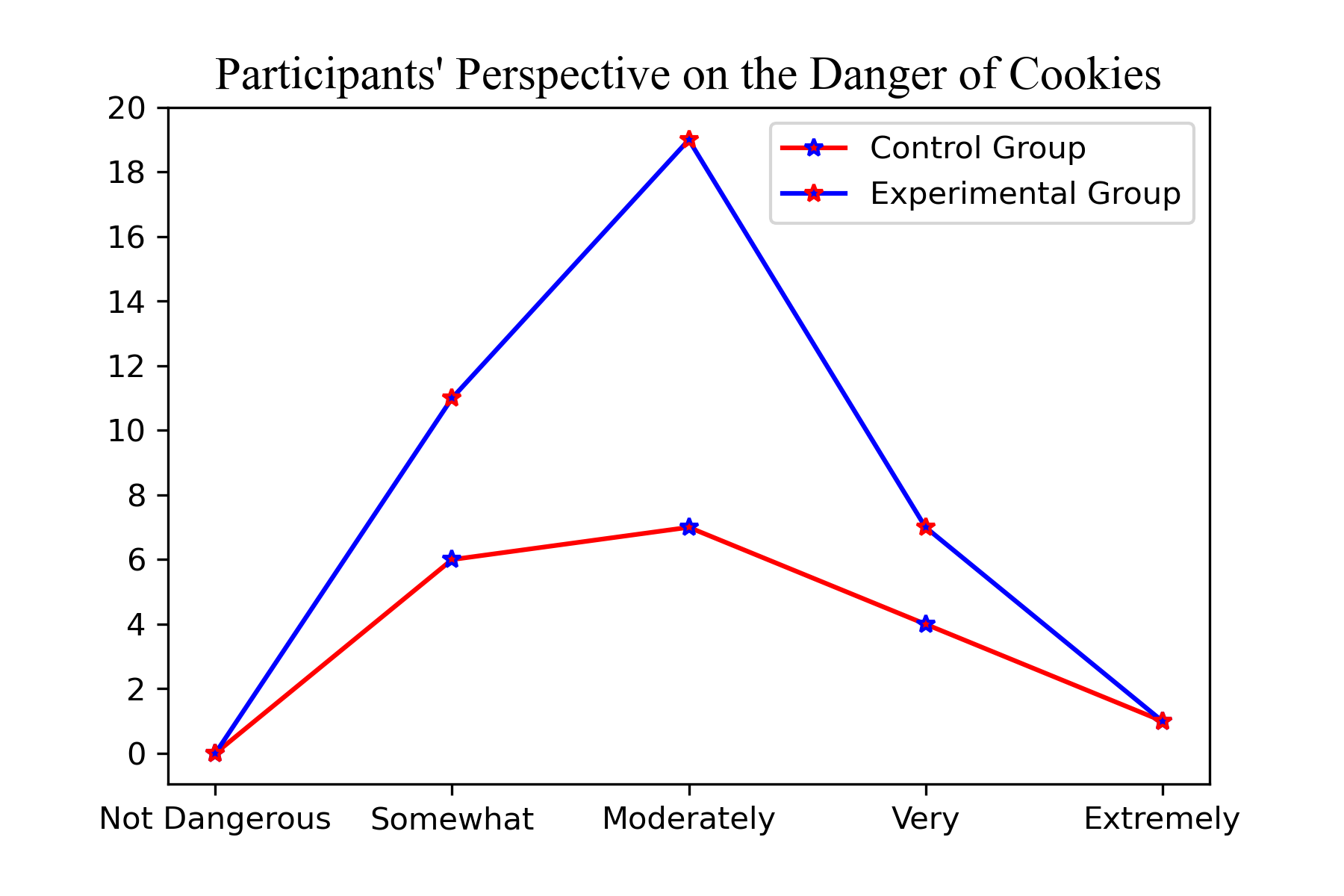}
\caption{\label{fig:5} Participants' perspectives on the danger of cookies in both groups}
\end{figure}

As mentioned, we had two groups; the control group ($34$), which was not provided with information, unlike the experimental group ($34$). All participants in the \textit{Control Group}  were transparent about the extent of their knowledge about cookies, as their choices said that 11.11\% do not know about cookies, and 33.33\% said they have little knowledge of websites' cookies. In contrast, the percentage of the \textit{Experimental Group} awareness about cookies increased clearly. The percentage of participants who answered that they did not know cookies was 0\%, while 41.38\% of the participants replied that they have as much as little knowledge, and 55.17\% were those with moderate Knowledge of website cookies meaning the educational materials supplied to the \textit{Experimental Group} had an impact on their overall knowledge. 
On the other hand, concerning threat perception of cookies, about 50\% of the \textit{Experimental Group} consider website cookies as a moderate threat to their online privacy, and 18.42\% consider cookies as very dangerous to their privacy and security. While in the \textit{Control Group}, 38.89\% believe cookies pose moderately tricky to their online privacy, and 33.33\% see cookies as somewhat risky. From figure~\ref{fig:5}, we could quickly point out that the educational materials used by the experimental group were able to provide them with more information about cookies. 
\section{Limitations and Future Work}
\label{sec:limit}
We conducted a risk-based analysis with $68$ participants to evaluate their perception of cookies and targeted advertisements. 
However, most of our participants belonged to the convenience sample of university students primarily from one geographical location. Though this study helped us conduct a mixed-methods analysis, we aim to achieve quantitative research to gauge the impact of risk communication. Thus, we plan to extend this survey to a broader audience via crowdsourcing platforms. In addition, we plan to implement eye trackers and other trackers in addition to the mouse trackers and attention checks in the future study to evaluate the impact of the learning models.
\section{Conclusion}
\label{sec:conclusion}
To study users' awareness about their keenness to preserve their online privacy while browsing websites, we conducted a user survey to identify the participants' browsing practices and how they deal with targeted advertisements and cookies with $68$ participants. We analyzed our research question from two sides using a control versus experiment group. The participants in the experiment group were shown more information about cookies and targeted advertisements through a website with textual data and informative videos. We developed the learning website through participatory design with $15$ individuals in a workshop. This bipartite group analysis helped us evaluate user risk perception and valuation through cybersecurity awareness and education. The experimental group with good knowledge about third parties and privacy risks reacted positively and realized the risks involved in publishing their personal information, and became more aware of the importance of tools to prevent cookies and targeted ads. Moreover, they desire to hide their address location in applications and websites. Hence, our study succeeds in providing important work toward spreading awareness about the dangers of cookies and third parties. 
\section{Acknowledgement}
We thank Hebah Alafari for the initial help with the study design and acknowledge the Inclusive Security and Privacy-focused Innovative Research in Information Technology (InSPIRIT) Lab at the University of Denver. Any opinions, findings, conclusions, or recommendations expressed in this experimental research are solely those of the authors.
\bibliographystyle{spmpsci.bst}
\bibliography{References}

\end{document}